\documentclass[runningheads]{llncs}
\usepackage{amsmath}
\usepackage{orcidlink}
\usepackage[T1]{fontenc}
\usepackage{graphicx}
\begin{document}
\title{QUBO formulations for NP-Hard spanning tree problems}
%
%
\author{Ivan Carvalho \orcidlink{0000-0002-8257-2103}}
\authorrunning{I. Carvalho}
%
\institute{University of British Columbia, Kelowna, Canada \\ \email{ivancarv@student.ubc.ca}}
\maketitle              
\begin{abstract}
We introduce a novel Quadratic Unconstrained Binary
Optimization (QUBO) formulation method for spanning tree problems. Instead of encoding the presence of edges in the tree
individually, we opt to encode spanning trees as a permutation problem. We apply our method to four NP-hard spanning tree variants, namely the \(k\)-minimum spanning tree,
degree-constrained minimum spanning tree, minimum leaf spanning
tree, and maximum leaf spanning tree.
Our main result is a formulation with \(\mathcal{O}(|V|k)\) variables
for the \(k\)-minimum spanning tree problem, beating related strategies that need \(\mathcal{O}(|V|^{2})\) variables.


\end{abstract}

\section{Introduction}

Given an edge-weighted undirected graph \(G = (V, E)\), the minimum
spanning tree (MST) of a graph is the subgraph \(G_{T} = (V, E_{T})\)
such that \(G_{T}\) is connected, acyclic, and the sum of the edge
weights, \(\sum_{(u, v) \in E_{T}} W_{u,v}\), is minimized. Finding the
MST of a graph can be done in polynomial time with greedy algorithms
such as Kruskal's algorithm \cite{Kruskal1956} and Prim's
algorithm \cite{Prim1957}.

However, adding constraints to the spanning tree or changing the
objective make some variants much harder to solve. Constraints such as
limiting the minimum number of vertices \cite{Ravi1996,Fischetti1994} or limiting the
maximum degree of vertex \cite{Ravi2001} make the problem
NP-Hard. The same applies to minimizing (or maximizing) the
number of leaves \cite{Salamon2008,Galbiati1994}.

In this paper, we focus on using Quadratic Unconstrained Binary
Optimization (QUBO) to solve NP-Hard spanning tree problems. We introduce a novel QUBO formulation method for spanning tree problems. Instead of encoding the presence of edges in the tree
individually, we opt to encode spanning trees as a permutation problem.
Our main result is a formulation with \(\mathcal{O}(|V|k)\) variables
for the \(k\)-minimum spanning tree problem (kMST). The method also yields a
formulation with \(\mathcal{O}(|V|^{2})\) variables for the
degree-constrained minimum spanning tree (DCMST), the minimum leaf spanning
tree, and the maximum leaf spanning tree.

\section{Quadratic Unconstrained Binary
Optimization}

QUBO instances are relevant because they can be approximately solved using methods such
as simulated annealing \cite{Kirkpatrick1983}, quantum
annealing \cite{Johnson2011}, digital annealing \cite{Aramon2019}, and
the Quantum Approximate Optimization Algorithm (QAOA) \cite{Farhi2014}.

Solving QUBO instances can be described as minimizing a quadratic
polynomial over binary variables \(x_{1}, x_{2}, \ldots, x_{n}\):

\[
{\arg\!\min}_{x \in \{ 0, 1 \}^{n} } \hspace{0.2cm} H = c + \sum_{i=1}^{n} \sum_{j=i+1}^{n} b_{i,j} x_{i}x_{j} + \sum_{i=1}^{n} a_{i} x_{i} 
\] QUBO formulations are also closely connected to Ising models, that
are described by a quadratic polynomial with variables
\(\sigma_{i} \in \{ -1, +1 \}\):

\[
{\arg\!\min}_{\sigma \in \{ -1, +1 \}^{n} } \hspace{0.2cm} H = - \sum_{i=1}^{n} \sum_{j= i + 1}^{n} J_{i,j} \sigma_{i} \sigma_{j} - \mu \sum_{i=1}^{n} h_{i} \sigma_{i}
\]

We highlight that there is a mapping between QUBO and Ising by using the
equation \(\sigma_{i} = 2x_{i} - 1\). Hence we can swap QUBO and Ising
formulations interchangeably. Ising formulations are more common in
Physics and are closer to the way hardware such as quantum annealers
represent the problem. QUBO formulations are more common in the field of
Computer Science, and will be the chosen notation for the remaining
equations.

There are two QUBO formulation metrics that are pertinent for the hardware that can solve the instances, namely the number of variables and the density of the formulation \cite{Fowler2017}. 

Existing hardware such as D-Wave's quantum annealers or Fujitsu's digital annealers can only solve QUBO instances up to a limited number of variables. Therefore, the best hardware-friendly formulations use the fewest possible number of variables.

Moreover, the number of interactions among the variables, called density, is also relevant to the hardware. D-Wave's quantum annealers do not have full connectivity among their qubits and often a minor-embedding \cite{Choi2010} is required to map the Ising model to the hardware topology. Minor-embeddings use a chain of qubits to represent a single variable, which reduces the size of the QUBO problems that can be solved by hardware.  Thus, the best hardware-friendly formulations also have the smallest possible density. 

\section{Related Work and Contributions}

The literature for QUBO formulations is plentiful, with given
formulations for many combinatorial optimization
problems \cite{Glover2022}. In his work, Lucas proposes QUBO
formulations for many NP problems, including the DCMST problem \cite{Lucas2014}. Lucas proposed
formulation uses \(\mathcal{O}(|V|^{3})\) variables in its worst case,
with the largest number of variables of the formulation being dedicated
to encode the depth of an edge \((u, v)\) in the spanning tree.

In his Master's thesis, Fowler improves upon Lucas' works and provides a
DCMST formulation that uses
\(\mathcal{O}(|V|^{2})\) variables in its worst
case \cite{Fowler2017}. Fowler's improved formulation relies on
encoding edges (which require \(\mathcal{O}(|E|)\) variables) and
ordering variable among the vertices (which require
\(\mathcal{O}(|V|^{2})\) variables).

Silva et al.~also offer a QUBO formulation for another NP-Hard variant of the
spanning tree problem, the minimum loss spanning tree reconfiguration
problem \cite{Silva2021}. Their formulation diverges from that of
Lucas and Fowler because it assumes that the input grah is planar, which
is valid for the electrical networks they analyse. Nevertheless, their
formulation still encodes the presence of each vertex in the tree and
requires \(\mathcal{O}(|E|)\) variables.

\subsection{Our Contributions}

Our new proposed formulations, by contrast, do not encode the presence
of edges in the spanning tree. Instead, we frame the spanning tree
problem as a permutation problem, similarly to Lucas' formulation for
the Travelling Salesman Problem (TSP) \cite{Lucas2014} .This novel
technique allows us to encode the \(k\)-minimum spanning tree problem
with \(\mathcal{O}(|V|k)\) variables, which uses fewer variables than
the \(\mathcal{O}(|V|^{2})\) related formulations for small \(k\). Using
fewer varaibles is desirable, as by the hardware limitations discussion earlier.

Another consequence of adopting the permutation strategy is that advancements in encoding permutations, generally focused on solving the TSP, will also benefit our encoding for spanning trees. Techniques such as using Higher Order Binary Optimization for QAOA \cite{Glos2022} or leveraging qudits instead of qubits \cite{Vargas2021} could also benefit our proposed encoding.

Our work is the first in literature to introduce QUBO formulations for the kMST and for the minimum (maximum) leaf spanning tree. Nonetheless, we believe that Fowler's DCMST formulation could encode those problems with minor tweaks while keeping \(\mathcal{O}(|V|^{2})\) variables. Hence, for comparison purposes, we refer to Fowler's work as a related strategy with \(\mathcal{O}(|V|^{2})\) variables. 

\section{Methods}

Our method is based on incrementally building a spanning tree by adding one vertex at a time. Throughout the building process, the spanning tree always remains a single connected component. Assume that the vertices are labelled \(1, 2, \ldots, |V| \). Then, the order that the vertices are added in the tree can be treated as a permutation of \( \{1, 2, \ldots, |V| \} \).

Hence, the first type of variable we use in our encoding
\(x_{v, i}\) is used to represent the positions of vertices in the permutation (similar to the QUBO encoding of the TSP):

\[
x_{v, i} = \begin{cases}
1, \hspace{0.5cm} \text{if the vertex } v \text{ is at the } i \text{-th position} \\
0, \hspace{0.5cm} \text{otherwise}
\end{cases}
\]

The method, however, also needs additional variables. One key difference between a spanning tree and a path is the edge used to connect a vertex \(v\). On a path, it is implied that if the vertex \(u\) is at position \(i\) and the vertex \(v\) is at position \(i+1\), then the edge \( (u,v) \) is used to connect the two. The same does not apply to spanning trees, because a vertex in the tree might have degree greater than two.

Thus, the second type of variable we use in our encoding \(y_{v, i}\) is used to represent which previous position in the permutation each vertex connects to:

\[
y_{v, i} = \begin{cases}
1, \hspace{0.5cm} \text{if the vertex } v \text{ has an edge connecting it to the vertex at the } i \text{-th position} \\
0, \hspace{0.5cm} \text{otherwise}
\end{cases}
\]

This representation is convenient for two reasons. Firstly, it let us enforce that a vertex only connects to previous vertices in the sequence. We can add a constraint that \( \sum_{i=1}^{|V|} \sum_{j=i}^{|V|-1} x_{v,i} y_{v,j} = 0  \) for each vertex \(v\). Secondly, we can test if an edge \( (u,v) \) is present in the tree with the expression \( \sum_{i=1}^{|V-1|} x_{u, i} y_{v,i}\).

\section{Results}

\subsection{\(k\)-Minimum Spanning Tree}

\subsubsection{Preliminaries}

Given an edge-weighted undirected graph \(G = (V, E)\) with non-negative edge weights and a positive integer \(k \leq |V|\), the kMST of a graph can be defined as the MST with at least \(k\) vertices. That is, the kMST is the subgraph \(G_{T} = (V_{T}, E_{T})\)
such that \(G_{T}\) is connected and acyclic, the number of vertices \(|V_{T}| \) is greater than or equal to \(k\), and the sum of the edge
weights, \(\sum_{(u, v) \in E_{T}} W_{u,v}\), is minimized. The kMST problem is also known as edge-weighted \(k\)-cardinality tree in literature \cite{Fischetti1994}. 

For the QUBO formulation, we will exploit the fact that the solution for the kMST has exactly \(k\) vertices \cite{Ravi1996}. For each vertex \(v \in V\), we define variables  \(x_{v, i}\) with \(1 \leq i \leq k\) for encoding the presence of the vertex in the permutation of size \(k\). We also define for each \(v \in V\) the variables \(y_{v,i}\) with \(1 \leq i \leq k - 1\) to encode the connections to previous vertices in the permutation. The formulation uses \(2 |V|k - |V|\) variables in total.

Our formulation also contains two parameters \(\lambda_{A} > 0\) and \(\lambda_{B} > 0\) to tweak the energy of the hamiltonian. We use \(\lambda_{A}\) as a coefficient for the quadratic penalties \cite{Glover2022}. QUBO problems are by definition unconstrained, hence we add penalties such that invalid solutions cannot be the global minimum of the expression.

\subsubsection{Formulation}

We formulate the Hamiltonian for the kMST using three parts such that \(H_{kMST} = H_{Tree} + H_{Cst}^{(1)} + H_{Cst}^{(2)}\).

The first Hamiltonian represents the cost of the edges in the kMST, by checking the presence of edges in the tree using the \( \sum_{i=1}^{k} x_{u, i} y_{v,i}\) expression discussed earlier. We also add a penalty in the Hamiltonian such that no invalid edge \( (u, v) \notin E\) is allowed in a solution to the QUBO.

\[
H_{Tree} = \lambda_{B}\sum_{(u, v) \in E} W_{u,v} \sum_{i = 1}^{k-1} x_{u, i} \cdot y_{v, i} + \lambda_{A}\sum_{(u, v) \notin E} \sum_{i = 1}^{k-1} x_{u, i} \cdot y_{v, i}
\]

The second Hamiltonian ensures that three constraints are met. The first constraint is that no two vertices must occupy the same position in the permutation. The second constraint is that no vertex may occupy two positions in the permutation. The third constraint is that for each position \(i \in \{1, 2, \ldots k\} \), exactly one vertex must occupy such position.

\[
H_{Cst}^{(1)} = \lambda_{A} \sum_{v=1}^{|V|} \sum_{u=v+1}^{|V|} \sum_{i=1}^{k} x_{u, i} \cdot x_{v,i} + \lambda_{A} \sum_{i=1}^{k} \sum_{j=i+1}^{k} \sum_{v=1}^{|V|} x_{v, i} \cdot x_{v,j}  + \lambda_{A} \sum_{i=1}^{k} (1 - \sum_{v=1}^{|V|}x_{v, i})^{2}
\]

The third Hamiltonian validates another two constraints related to connections of each vertex. The first constraint penalizes connections from a vertex to another vertex ahead of it in the permutation. The second constraint validates that if a vertex \(v\) is present in the permutation and it is not the first vertex, then exactly one of its \(y_{v,i}\) is set to \(1\).

\[
H_{Cst}^{(2)} = \lambda_{A} \sum_{u=1}^{|V|} \sum_{i=2}^{k} (x_{u, i} \cdot \sum_{j = i}^{k-1}y_{u, j} ) +  \lambda_{A} \sum_{v=1}^{|V|}(\sum_{i=2}^{k}x_{v,i} - \sum_{i=1}^{k-1}y_{v, i})^{2}
\]

Therefore, the Hamiltonian for the kMST is:

\[
H_{kMST} = H_{Tree} + H_{Cst}^{(1)} + H_{Cst}^{(2)}
\]

To ensure that the solution that minimizes the energy of the Hamiltonian is a valid solution to the kMST, we propose a lower bound to the penalty coefficient of \(\lambda_{A} > \lambda_{B} |V| \cdot \max_{(u, v) \in E} (W_{u,v}) > 0\).

The proposed formulation uses \(\mathcal{O}(|V|k)\) variables and contains \(\mathcal{O}(|V|^{2}k + |V|k^{2})\) interactions among the variables.

\subsection{Degree restricted Minimum Spanning
Tree}

\subsubsection{Preliminaries}

Given an edge-weighted undirected graph \(G = (V, E)\) with non-
negative edge weights and a positive integer \(\Delta\), the DCMST  of a graph can be defined as finding a MST such that the degree of every vertex is less than or equal to \(\Delta\). That is, the DCMST is the subgraph \(G_{T} = (V, E_{T})\)
such that \(G_{T}\) is connected and acyclic, \(\deg(v) \leq \Delta\) holds for every \(v \in V\), and the sum of the edge
weights, \(\sum_{(u, v) \in E_{T}} W_{u,v}\), is minimized.

The QUBO formulation for the DCMST is similar to the one for the kMST. The parameters \(\lambda_{A} > 0\) and \(\lambda_{B} > 0\) return on the formulation, and many of the Hamiltonians are similar. However, there are two key differences. The first difference is that the permutation is of size \(|V|\), hence we define variables \(x_{v,i}\) for each \(v \in V\) and for each \(i\) with \(1 \leq i \leq |V|\). Analogously, we also define \(y_{v, i}\) for each \(v \in V\) and for each \(i\) with \(1 \leq i \leq |V| - 1\).

The second difference is the use of degree-counter slack variables for encoding the inequality constraints. We use an identical strategy to the one in Lucas' and Fowler's works \cite{Lucas2014,Fowler2017}. Let \(M = \lfloor \log_{2} \Delta \rfloor \). Then, for each position \(i \in \{1, 2, \ldots |V| - 1\}\) in the permutation, we define the expression \(\mathcal{Z}_{i}\) that counts how many vertices connect to the position \(i\): 

\[ \mathcal{Z}_{i} = (\sum_{j=0}^{M-1} z_{i,j} 2^{j}) + z_{i,M}(\Delta + 1 - 2^{M})  \]

Each variable \(z_{i,j}\) can be interpreted as a bit in the binary representation of \(\mathcal{Z}_{i}\), with the exception of \(z_{i,M}\) which is a remainder for when \(\Delta\) is not of the form \(2^{n} - 1\). We point that \(0 \leq \mathcal{Z}_{i} \leq \Delta\), hence the degree-counter variable cannot assume values that are not consistent with a DCMST solution.

The formulation uses \( 2|V|^2 + |V|\lfloor \log_{2} \Delta \rfloor - ( \lfloor \log_{2} \Delta \rfloor + 1)  \) variables in total.

\subsubsection{Formulation}

The Hamiltonian for the DCMST is formulated with four parts such that \(H_{DCMST} = H_{Tree} + H_{Cst}^{(1)} + H_{Cst}^{(2)} + H_{Cst}^{(3)}\).

The first Hamiltonian represents the cost of the edges in the DCMST, by checking the presence of edges in the tree. Likewise, there is a penalty in the Hamiltonian for invalid edges. \(H_{Tree}\) is almost identical to the one for the kMST, with the difference being on the sum of the terms going from \(1\) to \(|V|\) instead of \(1\) to \(k\).

\[
H_{Tree} = \lambda_{B}\sum_{(u, v) \in E} W_{u,v} \sum_{i = 1}^{|V|-1} x_{u, i} \cdot y_{v, i} + \lambda_{A}\sum_{(u, v) \notin E} \sum_{i = 1}^{|V|-1} x_{u, i} \cdot y_{v, i}
\]

The second Hamiltonian checks constraints regarding the permutation. The first constraint is that each vertex must appear exactly once in the permutation. The second constraint is that for each position, exactly one vertex must be assigned to it.

\[
H_{Cst}^{(1)} = \lambda_{A} \sum_{v=1}^{|V|} (1 - \sum_{i=1}^{|V|}x_{v, i})^{2} + \lambda_{A} \sum_{i=1}^{|V|} (1 - \sum_{v=1}^{|V|}x_{v, i})^{2}
\]

The third Hamiltonian checks two other constraints concerning the connection of the nodes. The first constraint penalizes connections from a vertex to another vertex ahead of it in the permutation. The second constraint ensures that every vertex connects to exactly one previous vertex, with the exception of the first vertex does not require such connection.

\[
H_{Cst}^{(2)} =  \lambda_{A} \sum_{u=1}^{|V|} \sum_{i=2}^{|V|} (x_{u, i} \cdot \sum_{j = i}^{|V|-1}y_{u, j} ) + \lambda_{A} \sum_{v=1}^{|V|}(1 - x_{v, 1} - \sum_{i=1}^{|V|-1}y_{v, i})^{2}
\]

The fourth Hamiltonian is for the degree related constraints. For each \(i\), we validate that \( \mathcal{Z}_{i} \) and its associated \(z_{i,j}\) variables encode the number of connections of position \(i\). Note that for \(i \geq 2\), we also need to subtract one to account for the implicit connection with a previous vertex in the permutation.

\[
H_{Cst}^{(3)} = \lambda_{A} (\mathcal{Z}_{1} - \sum_{v=1}^{|V|}y_{v,1})^{2} + \lambda_{A} \sum_{i=2}^{|V|-1} (\mathcal{Z}_{i} - 1 - \sum_{v=1}^{|V|}y_{v,i})^{2}
\]

Therefore, the Hamiltonian for the DCMST is:

\[
H_{DCMST} = H_{Tree} + H_{Cst}^{(1)} + H_{Cst}^{(2)} + H_{Cst}^{(3)}
\]

We propose again the lower bound for the penalty coefficient of \( \lambda_{A} > \lambda_{B} |V| \cdot \max_{(u, v) \in E} (W_{u,v}) > 0\), to ensure the ground state of the Hamiltonian is a valid DCMST solution.

The proposed formulation uses \(\mathcal{O}(|V|^{2})\) variables and contains \(\mathcal{O}(|V|^{3})\) interactions among the variables.

\subsection{Minimum (Maximum) Leaf Spanning Tree}

\subsubsection{Preliminaries}

Given a unweighted undirected graph \(G = (V, E)\), the minimum leaf spanning tree (MLST)   of a graph can be defined as the spanning tree such that the number of leaves is minimized. Let \(V_{L} = \{v \in V | \deg(v) = 1  \}\) denote the leaves of the tree. Then, the MLST is the subgraph \(G_{T} = (V, E_{T})\)
such that \(G_{T}\) is connected and acyclic, and the number of leaves \(|V_{L}|\) is minimized.

A related problem to the MLST is the maximum leaf spanning tree, which maximizes the number of leaves \(|V_{L}|\). We will show that the QUBO formulation for the two problems is almost identical, as we can swap minimizing for maximizing by multiplying by minus one. For the formulation discussion, we focus on the problem that minimizes leaves for simplicity.

The QUBO formulation for the MLST shares many similarities with the one for the kMST. The formulation uses the variables \(x_{v,i}\), \(y_{v, i}\), and the parameters \(\lambda_{A} > 0\) and \(\lambda_{B} > 0\) just like the DCMST. The major difference is on the slack variables, which instead of just counting the degree also check if a vertex is a leaf in the tree.

We use a technique similar to the idea from the Max \(k\)-SAT QUBO formulation \cite{Nlein2022} to count leaves in the tree. We add a depth \(d\) to the degree-counter expressions \(\mathcal{Z}_{d,i}\) . For \(d = 1\), \(\mathcal{Z}_{1,i}\)  counts \(\deg(v) - 1\), where \(v\) is the vertex such that \(x_{v,i} = 1\). For \(d > 1\), \(\mathcal{Z}_{d,i}\) counts how many variables were set to one in \(\mathcal{Z}_{d-1,i}\). For some \(D = d\), \(\mathcal{Z}_{D,i}\) will eventually be represented by exactly two binary variables. 

Because \(\deg(v) - 1\) is equal to zero if \(v\) is a leaf, \(\mathcal{Z}_{1,i}\) has no variable set to one if the vertex at position \(i\) is a leaf. This propagates to the next depths, as \(\mathcal{Z}_{d,i}\) will also have no variable set to one. Therefore, we can verify that the vertex at position \(i\) is a leaf by checking if \(\mathcal{Z}_{D,i} = 0\).

Let \(D = \log_{2}^{\star} |V|\) be the iterated logarithm of the number of vertices. We define a sequence \(M\) such that \(M_{0} = |V| - 1\) and \(M_{d} =  \lfloor \log_{2} (1 + M_{d-1}) \rfloor \) for \(1 \leq d \leq D\). Then, we define the expression \(\mathcal{Z}_{d,i}\) for each position \( i \in \{1, 2, \ldots , |V| - 1\}\) and each depth \(1 \leq d \leq D\): 

\[ \mathcal{Z}_{d,i} = (\sum_{j=0}^{-1 + M_{d}} z_{d,i,j} 2^{j}) + z_{d,i,M_{d}}(M_{d-1} + 1 - 2^{M_{d}})\]

Each variable \(z_{d,i,j}\) can be interpreted as a bit in the binary representation of \(\mathcal{Z}_{d,i}\), with the exception of \(z_{d,i,M_{d}}\) which is a remainder for when \(M_{d}\) is not of the form \(2^{n} - 1\).

For convenience, we also define the expression \(\mathcal{B}_{d,i}\) that counts the number of variables that are set in \(\mathcal{Z}_{d,i}\):

\[ \mathcal{B}_{d,i} = \sum_{j=0}^{M_{d}} z_{d,i,j}  \]

Let \(L = \sum_{d=1}^{D} 1 + M_{d}\) denote the sum of slack variables associated with a position. Then, the total number of variables for the formulation is \(2|V|^{2} - |V| + L|V| - L \).

\subsubsection{Formulation}

The Hamiltonian for the MLST is formulated with five parts such that \(H_{MLST} = H_{Tree} + H_{Edges} + H_{Cst}^{(1)} + H_{Cst}^{(2)} + H_{Cst}^{(3)}\).

The first Hamiltonian counts the number of leaves. The expression for counting leaves \((1 - z_{D,i, 1})(1 - z_{D,i,0})\) evaluates to one if the vertex at position \(i\) is a leaf. Otherwise, it evaluates to zero as at least one of the bits will be set. Notice that the vertex at position \(i = |V|\) is always a leaf, hence we add a constant to account for that.

\[
H_{Tree} = \lambda_{B} ( 1 + \sum_{i=1}^{|V| - 1} (1 - z_{D,i, 1})(1 - z_{D,i,0}) )
\]

The second Hamiltonian is for penalizing invalid edges \((u, v) \notin E\).

\[
H_{Edges} = \lambda_{A}\sum_{(u, v) \notin E} \sum_{i = 1}^{|V|-1} x_{u, i} \cdot y_{v, i}
\]

The third Hamiltonian checks constraints regarding the permutation, and is identical to \(H_{Cst}^{(1)}\) from the DCMST formulation. 
\[
H_{Cst}^{(1)} = \lambda_{A} \sum_{v=1}^{|V|} (1 - \sum_{i=1}^{|V|}x_{v, i})^{2} + \lambda_{A} \sum_{i=1}^{|V|} (1 - \sum_{v=1}^{|V|}x_{v, i})^{2}
\]

The fourth Hamiltonian checks constraints concerning the connection of the nodes, and is identical to \(H_{Cst}^{(2)}\) from the DCMST formulation.

\[
H_{Cst}^{(2)} =  \lambda_{A} \sum_{u=1}^{|V|} \sum_{i=2}^{|V|} (x_{u, i} \cdot \sum_{j = i}^{|V|}y_{u, j} ) + \lambda_{A} \sum_{v=1}^{|V|}(1 - x_{v, 1} - \sum_{i=1}^{|V|}y_{v, i})^{2}
\]

The fifth Hamiltonian is for the constraints related to \(\mathcal{Z}_{d,i}\). For \(d = 1\), we ensure that \(\mathcal{Z}_{1,i}\) is equal to the degree minus one. The expression for \(i = 1\) is slightly different than the one for \(i \geq 2\) because the latter ones have an implicit connection to a previous position in the permutation. For \(2 \leq d \leq D\), we verify the constraint that \(\mathcal{Z}_{d,i} = \mathcal{B}_{d-1,i}\).

\[
H_{Cst}^{(3)} = \lambda_{A} ( (\mathcal{Z}_{1,1} + 1 - \sum_{v=1}^{|V|} y_{v,1} )^{2} + \sum_{i=2}^{|V| - 1}(\mathcal{Z}_{1,i} - \sum_{v=1}^{|V|} y_{v,i} )^{2} ) + \lambda_{A} \sum_{d=2}^{D} \sum_{i=1}^{|V|- 1} (\mathcal{Z}_{d,i} - \mathcal{B}_{d-1,i})^{2}
\]

Therefore, the Hamiltonian for the MLST is:

\[
H_{MLST} = H_{Tree} + H_{Edges} + H_{Cst}^{(1)} + H_{Cst}^{(2)} + H_{Cst}^{(3)}
\]

We suggest the lower bound for the penalty coefficient of \( \lambda_{A} > \lambda_{B} |V| > 0\), to ensure the ground state of the Hamiltonian is a valid MLST solution.

To solve for the maximum leaf spanning tree, we just swap the sign of \(H_{Tree}\) to \(-H_{Tree}\). The sign swap rewards leaves in the tree instead of adding cost, hence the number of leaves is maximized.

\[
H_{MaxLeaf} = -H_{Tree} + H_{Edges} + H_{Cst}^{(1)} + H_{Cst}^{(2)} + H_{Cst}^{(3)}
\]

The proposed formulations use \(\mathcal{O}(|V|^{2})\) variables and contain \(\mathcal{O}(|V|^{3})\) interactions among the variables.

\section{Conclusion}

We have provided a novel method to encode NP-Hard spanning tree problems as QUBO instances. We leveraged the well-known permutation problem method to encode the TSP as a QUBO, and tweaked it to apply it to spanning trees.

For the kMST, our novel formulation of \(\mathcal{O}(|V|k)\) variables yields an improvement in the number of variables required to encode the problem compared to related strategies that require \(\mathcal{O}(|V|^{2})\) variables. For the DCMST and for the MLST, our novel formulations use \(\mathcal{O}(|V|^{2})\) variables and match related encodings with regards to the number of variables.

Future work on this topic is an invitation to transfer recent improvements from QUBO research of the TSP into spanning tree problems. In particular, the use of Higher Order Binary Optimization seems especially promising as it reduces the number of binary variables and qubits to encode spanning tree instances.

%
%
%
\bibliographystyle{splncs04}
\bibliography{mybibliography}

\end{document}